# I-Min: An Intelligent Fermat Point Based Energy Efficient Geographic Packet Forwarding Technique for Wireless Sensor and Ad Hoc Networks

Kaushik Ghosh[1], Sarbani Roy [2] & Pradip K. Das [3]

[1,3] Mody Institute of Technology & Science , Lakshmangargh,(Sikar) India
k79aushik@gmail.com , pkdas@ieee.org
[2] Jadavpur University, Kolkata, India
sarbani.roy@gmail.com

## ABSTRACT

*Energy consumption and delay incurred in packet delivery are the two important metrics for measuring the performance of geographic routing protocols for Wireless Adhoc and Sensor Networks (WASN). A protocol capable of ensuring both lesser energy consumption and experiencing lesser delay in packet delivery is thus suitable for networks which are delay sensitive and energy hungry at the same time. Thus a smart packet forwarding technique addressing both the issues is thus the one looked for by any geographic routing protocol. In the present paper we have proposed a Fermat point based forwarding technique which reduces the delay experienced during packet delivery as well as the energy consumed for transmission and reception of data packets.*

## KEYWORDS

*Geographic routing, Fermat point, Wireless Adhoc and Sensor Network(WASN).*

## 1. INTRODUCTION

Geographic routing protocols are fast coming up in the arena of Wireless Adhoc and Sensor Networks (WASN) for obvious reasons. Mobile Adhoc Networks (MANET) are another area where geographic routing protocols find their application. This is because, adhoc and sensor networks are more and more used for applications where manual collection and assimilation of data is either troublesome or is impossible. Eg military surveillance, collecting temperature and humidity of uninhabited regions etc. In scenarios like this, data packets may need to be transmitted to one or more favorable locations where final computations on those data can take place. So, the possibility of the final destination node(s) to reside in a specific geographic region(s) for a mobile sensor network can by no means be ruled out. Now if the situation demands the network either to be deployed for a prolonged period of time or if the frequency of data transfer in the said network is considerably high, then the participating nodes in the network need to have a sound reserve of residual energy. Needless to say that replenishing the battery time and again in the networks under discussion either adds up non-computation/non-communication overhead or may turn out to be an impossible proposal altogether. The only way to retain uninterrupted service in such a scenario is to design a routing protocol that consumes the least possible energy. One way to reduce the transmitting/receiving energy of the network is to reduce the frequency of packet transmission without compromising the reliability. This can be done by waiting for a considerable number of packets to a particular destination to arrive first. A number of short transmissions are replaced by a single long transmission. But this approach is sure to take a toll on the performance of delay sensitive networks.

Achieving energy efficiency and delay reduction in a single protocol is possible when one or more parameters directly proportional to both can be reduced. The transmitting distance can be considered as such a parameter. Energy consumed for transmitting a packet is a directly proportional to the square of the distance between the sender and receiver along with being dependent on the size of the packet transmitted/received [6][7]. Assuming the transmission radii of all the nodes in the network to be equal, it can be said that lesser the distance between two nodes, lesser is the number of hops required for packet transmission and thus lesser is the delay. Because, the delay experienced in packet delivery is a function of the processing time of the intermediate nodes taken together. Considering the packet size for a said network using a specific routing protocol to be constant, the energy consumption along with delay in a particular network can thus be reduced by reducing the transmission path length. Now for a specific geographic region, where the final destination(s) is(are) fixed or changes seldom, position of remaining nodes plays a vital role in determining the energy consumed and delay experienced by a network. But in most of the real life applications the nodes are deployed in a random manner. Thus to minimize the total transmitting distance for a multi-destination network, one needs to find out the minimum of the sum of the distances from a particular source to all the destinations. If one considers the source node and the destinations to be the vertex of a triangle or a polygon, then the sum of the distance of the source to the Fermat point [1] and the distances from the Fermat point to all the destinations of the considered triangular/polygonal region is always minimum.

In this paper we have proposed an energy efficient routing protocol that works on the principle of Fermat point. In [2] and [3] the authors have proposed Fermat point based geographic routing mainly for two geocast regions. They have used a geometric method of finding the Fermat point. In [4] however a global minima based scheme was proposed by the authors to find the Fermat point for n geocast regions which out performed the geometry driven scheme discussed in [3] in terms of total distance traveled by a data packet and the energy thus consumed by the network in trafficking the packet from source to the destined geographic regions. The forwarding technique used in both [3] and [4] is greedy in nature. The present scheme uses the 'Global Minima' based scheme used in [4] for finding the Fermat point and the forwarding technique is also greedy in genre. However it makes certain intelligent improvisations in the forwarding technique used in [4] to get over the unnecessary route looping that was present in both [3] and [4]. As the proposed protocol is free from route looping, it further reduces the effective distance traveled and the total number of hops encountered by a packet. This way, the energy expenditure for the network gets reduced as well.

The rest of the paper is organized as follows: In section 2 we have discussed some relevant related works. Section 3 discusses the present forwarding technique and compares it with the schemes discussed in [3] and [4]. In section 4 we have the results and section 5 is the concluding section of this paper with a direction for possible future works.

## 2. RELATED WORKS

Geographic routing protocols of the present day have evolved a long from the days of Location Aided Routing (LAR) [9] and Location Based Multicast (LBM) Algorithms[8] and have taken Euclidean geometry into its stride. In LAR the authors introduced the concepts of *expected zone* and *request zone* in order to cut down the total number of transmissions as compared to simple flooding. LBM reduced the total number of transmissions by introducing the concept of *forwarding zone* , where not all the neighbors of a forwarding node were considered as intermediate hops. Instead a potential region was selected as the forwarding zone and nodes residing within that region were considered as intermediate nodes on the way to the multicast region. In [18] however a random selection of the relaying node was made via contention among receivers. Authors in [16] compared the effect of cooperative communication to direct transmission for different set of distances in wireless sensor networks. They concluded that for small distance between the source and destination, direct transmission is more energy efficient than cooperative communication.

Reducing the number of transmissions automatically reduces the energy consumption. Although technically there can be multiple multicast regions but authors in [3] were first to introduce the concept of multiple destination regions (geocast groups) for that matter. There they formed a triangular region with two geocast regions and the source as three vertex of the triangle. The Fermat point for the triangle was found using a **geometry driven scheme** [3] and the packet from the source to both the destination was ferried via the Fermat point. This way the minimum possible path length was selected for a certain deployment fashion of the nodes. Because, transmission energy required for wireless communication increases superlinerly with the communication distance[17]; a fact well supported in [4]. In case if there is no node situated in the Fermat point thus spotted, the node nearest to the point is selected. The packet forwarding technique used here is greedy in nature. The scheme worked fine for a triangular region but as the number of geocast region went on exceeding from two, the degree of accuracy in spotting the Fermat point went on decreasing while using the heuristic scheme proposed in [3] for spotting the Fermat point of a polygonal region. As a result, the transmission path length from the source to n geocast regions is no longer the minimum one possible. This makes the whole exercise of finding the Fermat point for a polygonal region as well as the motivation behind it meaningless. This drawback was addressed in [4] where the authors adopted a *minima* based approach to spot the Fermat point for a triangular/polygonal region. In the minima based scheme, one always ended up with the minimum possible path length from the source to destination via the Fermat point for a certain node deployment pattern.

The concept of Fermat point has been exploited till a considerable extent in [2][3] and [4]. But neither [3] or [4] could get rid of the redundant route looping that generates in the process of greedy forwarding [10] of the packets. In this paper we thus propose a scheme which works on the platform provided by the scheme in [4] and outscores [4] when it comes to reduced cases of route looping. In fact the present scheme is almost always free from route looping. As a result the consumption of energy while following the present scheme is even lesser when compared with [4] .

## 3. PROPOSED I-MIN SCHEME

The proposed scheme uses [4] as the platform to achieve minimum energy consumption while transmitting a **m** bit packet from source to **n** geocast regions via the Fermat point. Just like [4] the proposed scheme also uses a global minima (figure 1) based scheme for locating the Fermat point. In the present scheme, before every packet forwarding, a check is being made to see whether the destination node is reached or not. In case the destination node happens to be one of the neighbors of the the node presently holding the packet, the packet is simply unicasted instead of following the greedy method of forwarding. This checking was however absent in both [3] and [4]. This actually reduces route looping up to a considerable extent which in turn eliminates redundant forwarding to reduce the energy consumption. With the reduction in route looping, the number of hops encountered by a packet also reduces which in turn is sure to reduce the delay in packet delivery along with the reduction in energy consumption. For example let us consider the scenario depicted in figure 1. Let us consider node 1 to be possessing the packet for the time being. The dashed circle is the transmitting radius of node 1. If node 2 is the destination node, then following a simple greedy forwarding scheme would make 1 to forward the packet to node 3. This leads to the consumption of an unnecessary energy. Moreover in this type of a situation infinite route looping may also generate thus consuming a lot of useless energy when a simple path from 1 to 2 is very much present. The forwarding algorithm for the present scheme is given in figure 3.

The results in the next section shows that this I-Min scheme guarantees to be a minimum energy consuming packet forwarding scheme. It scores over its predecessor schemes of [3] and [4] in terms delay experienced in packet receipt as well.

## Minima Algorithm

Input: Coordinates of the sender node and that of different geocast regions.
Output: (fx, fy); Coordinates of the Fermat_Point.
Tdist : Total distance traveled by the packet.
Tpow: The sum of power consumed by all the intermediate nodes to forward the packet to m geocast regions

```
1.    max_x=MAX_x (Sx, GRX(N));
2.    max_y=MAX_y (Sy, GRY(N));
3.    min_x=MIN_x (Sx, GRX(N));
4.    min_y=MIN_y (Sx, GRY(N));
5.    dx=0;    /*          Initialize dx */
6.    dy=0     /* Initialize dy */
7.    flag=0; /* To check the Fermat Point*/
8.    for ( i=min_x; i<max_x; i++ )
9.    {  if (flag==1) break; /*  Fermat point found  */
10.   for ( j=min_y; j=max_y; j++ )
11.   { x=i;
12.   y=j;
13.   for ( k=0;k<n;k++ )
14.   { dx+= termdx ( x,GRX(k),y,GRY(k) );
15.   dy+= termdy ( y,GRY(k),x,GRX(k) );
16.   } /* end of for loop (line-11)*/
17.   if  ( dx==0 && dy==0 )
18.   { flag=1; /* Fermat point found */
19.   break;
20.   }
21.   dX=0; dY=0;
22.   } /* end of for loop (line-8) */
23.   }/* end of for loop (line – 6) */
24.   if ( flag==1 )
25.   { fx=x; fy=y; } /* Fermat point */
26.   Tdist= Total_Dist ( Sx, Sy, GRX(N), GRY(N), fx, fy );
27.   Tpow= Total_Pow ( Tdist )
```

Figure 1. Minima Algorithm for finding the Fermat point.

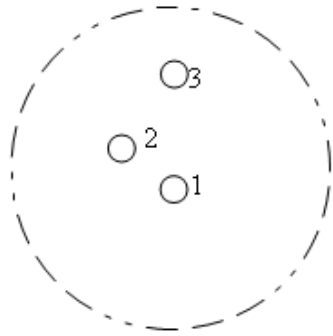

Figure 2. Selecting next hop neighbor when the destination node is within the transmission range of a forwarding node.

Table 1. Variables used in I-Min algorithm.

| Symbols | Definitions |
|---|---|
| D_ID | Node id of the destination. |
| NN | Number of neighbors of a forwarding node. |
| Neigh_Id | Node id of a neighbor. |
| neigh[] | Array containing the node id of the neighboring nodes. |
| next_hop | Node id of the next hop. |

**I-Min Algorithm**

```
Input: Node_id of the destination.
Output: Boolean.
D_ID: Node_id of the destination.
NN: Number of neighbors of the forwarding
node.
Neigh_ID= Node_id of a neighbor.

1. if (flag==0)
2. next_hop=MAX_DIST(D_ID, NN, neigh[]);
3. else
4. next_hop=D_ID;
5. for(i=0; i<NN; i++)
6. {if (D_ID==Neigh_ID)
7. flag=1;
8. else
9. flag=0; }
```

Figure 3. I-Min forwarding algorithm.

## 4. RESULTS

For simulation environment we have considered a region of 1800x1100 square meters. The nodes are deployed in a pseudo-random fashion within that region. The performance metrics taken are i) total number of hops traveled by a packet via the Fermat point to different geocast regions and ii) energy consumed to move the packet to different geocast regions. Results in [4] have clearly shown that using the global minima scheme, reduces the total distance a packet travels when compared with the geometry driven scheme. As a result the energy consumed for forwarding a packet to n geocast regions via the Fermat point is also less when one uses the Global Minima scheme in place of the geometry driven scheme. I-Min technique further reduces the total distance traveled by a packet by eliminating the redundant transmissions/receptions that were present in [4] due to flat greedy forwarding technique.

The graphs 1 and 2 convincingly concludes that the present scheme performs better when it comes to energy consumption and delay. We have used the radio model followed in [6] and [7] for calculating the total energy consumption in the network.

Table 2. Function used in I-Min algorithm and its definition.

| Functions | Definitions |
|---|---|
| MAX_Dist (D_ID,NN,neigh[]) | **Input:** Node id of the destination, Number of neighbors, Node id of all the neighbors. |

The following two equations have been used in this paper respectively for energy consumed for transmitting (ETX) and receiving (ERX) a data packet comprising of m bits. The distance between the sender and receiver is d. ETX is a function of both m and d, whereas ERX is the function of m alone.

$$E_{TX}(m,d) = m*E + m*\epsilon*d^2$$

$$E_{RX}(m) = m*E$$

Where,

$E$= 50nJ/bit and $\varepsilon$= 10 pJ/bit /m².
$E_{TX}$=Energy consumed for transmission.
$E_{RX}$=Energy consumed for reception.
$d$=distance between the transmitting and receiving node.
$\varepsilon$=Permittivity of free space.
$m$=Number of bits.

Table 3. Variables used in Global Minima algorithm.

| Symbols | Definitions |
| --- | --- |
| max_x | Maximum value of X coordinates |
| max_y | Maximum value of Y coordinates |
| min_x | Minimum value of X coordinates |
| min_y | Maximum value of Y coordinates |
| (Sx, Sy) | Coordinates of the sender |
| n | Number of geocast regions |
| GRX[] | Array storing the values of X-coordinates of all the geocast regions |
| GRY[] | Array storing the values of Y-coordinates of all the geocast regions |
| dx | Minimum value of X |
| dy | Minimum value of Y |
| (fx, fy) | Coordinates of the fermat point |
| Tdist | Total distance of transmission |
| Tpow | Total power consumed to transmit a packet from the sender to all the geocast regions |

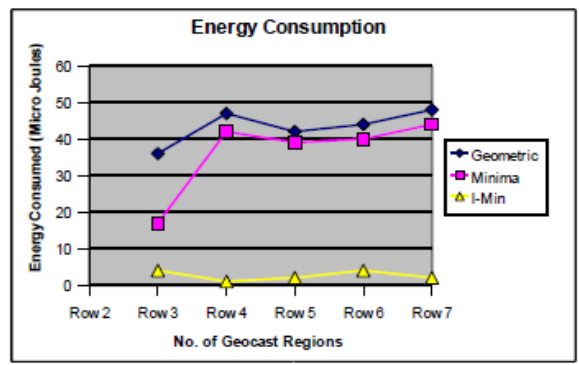

Graph 1. Energy consumption in the three different schemes.

Table 4. Functions used in Global Minima algorithm and their definitions.

| Functions | Definitions |
|---|---|
| `MAX_x(Sx, GRX[])` | **Input:** X coordinate of the sender and X coordinates of all the geocast regions. **Output:** Maximum value amongst the X coordinates. |
| `MAX_y (Sy,GRY[] )` | **Input:** Y coordinate of the sender and Y coordinates of all the geocast regions. **Output:** Maximum value amongst the Y coordinates. |
| `MIN_x(Sx, GRX[])` | **Input:** X coordinate of the sender and X coordinates of all the geocast regions. **Output:** Minimum value amongst the X coordinates. |
| `MIN_y(Sy, GRY[])` | **Input:** Y coordinate of the sender and Y coordinates of all the geocast regions. **Output:** Minimum value amongst the Y coordinates. |
| `termdx(x, GRX[], y,GRY[])` | $$\frac{\delta}{\delta x} f(x) \| y = constant$$ **Input:** X and Y coordinate of any point within the polygon and (X,Y) coordinates of all the geocast regions. **Output:** dx i.e., $\frac{\delta}{\delta x} f(x) \| y = constant$ |
| `termdy(y, GRY[], x,GRX[])` | $$\frac{\delta}{\delta y} f(y) \| x = constant$$ **Input:** X and Y coordinate of any point within the polygon and (X,Y) coordinates of all the geocast regions. **Output:** dy i.e., $\frac{\delta}{\delta y} f(y) \| x = constant$ |
| `Total_Dist(Sx, Sy, GRX[], GRY[], fx, fy)` | **Input:** (X,Y) Coordinates of the sender, (X,Y) coordinates of all the geocast regions and (X,Y) coordinates of the Fermat point. **Output:** Total distance from the sender to all the geocast regions via the Fermat point, taken individually. |
| `Total_Pow (Tdist)` | **Input:** Total distance of the path from sender to the Fermat point and from there to all the geocast regions taken individually. **Output:** Total power consumed for transmitting a packet to all the geocast regions. |

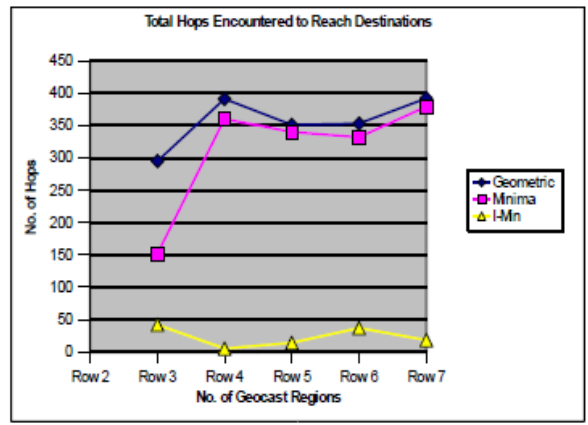

Graph 2. Comparison between three schemes with respect to number of hops encountered by a packet to travel from a given source to destination.

## 5. CONCLUSIONS

The results in the previous section clearly proves the energy friendly nature of the I-Min forwarding technique. In fact, any packet forwarding scheme capable of ensuring a lesser transmission distance is surely going to score over others when it comes to energy consumption, for that matter. A lesser number of hop counts again ensures lesser amount of delay for packet delivery. Because with an increase in the number of hops, the transmission time for packet delivery increases due to the increased total processing time. As future work, comparing the present scheme with other geographic routing techniques taking packet delivery ratio as the metric can be a possible option.

## REFERENCES

[1]     The Fermat point and generalizations. [Online]: http://www.cut-the-knot.org/generelization /fermat_point.shtml.

[2]     Young-Mi Song, Sung-Hee Lee and Young-Bae Ko"*FERMA: An Efficient Geocasting Protocol for Wireless Sensor Networks with Multiple Target Regions*" T. Enokido et al. (Eds.): EUC Workshops  2005, LNCS 3823, pp. 1138 – 1147, 2005. © IFIP International Federation for Information  Processing 2005.

[3]     S.H. Lee and Y.B. Ko. "*Geometry-driven Scheme forGeocast Routing in Mobile Adhoc Networks*", IEEE Transactions for Wireless Communications, vol.2, 06/ 2006.

[4]     Kaushik Ghosh, Sarbani Roy and Pradip K. Das, "*An Alternative Approach to find the Fermat Point  of a Polygonal Geographic Region for Energy Efficient Geocast Routing Protocols:Global Minima Scheme*", AIRCC/IEEE NetCoM 2009.

[5]     Lynn Choi, Jae Kyun Jung, Byong-Ha Cho and Hyohyun Choi, "*M-Geocast: Robust and Energy- Efficient Geometric Routing for Mobile Sensor Networks*", SEUS 2008, LNCS 5287,pp. 304–316, IFIP (International Federation for Information Processing) 2008.

[6]     W.B. Heinzelman, A.P. Chandrakasan and H. Balakrishnan, "*An application-specific protocol architecture for wireless micro sensor networks*", IEEE Transactions on Wireless Communications, Vol.1, Issue 4, October 2002.

[7]     I-Shyan Hwang and Wen-Hsin Pang, "*Energy Efficient Clustering Technique for Multicast Routing Protocol in Wireless Adhoc Networks*", IJCSNS, Vol.7, No.8, August 2007.

[8]     Young-Bae Ko and Nitin H. Vaidya "*Geocasting in Mobile Ad Hoc Networks: Location-Based Multicast Algorithms* ", Proceedings of the Second IEEE Workshop on Mobile Computer Systems and     Applications (WMCSA), February 1999.

[9]     Y.B. Ko and N. H. Vaidya, "*Location-aided routing (LAR) in Mobile Ad Hoc Networks*", ACM/Baltzer Wireless Networks (WINET) journal, vol. 6, no. 4, 2000, pp. 307-321.

[10]    B.Karp and H.Kung, "*GPSR: Greedy perimeter stateless routing for wireless networks*", ACM/IEEE MobiCom, August 2000.

[11]    C.Maihofer, "*A survey of geocast routing protocols*", IEEE Communications Survey & Tutorials,vol.6, no.2, Second Quarter 2004.

[12]    T Camp, Y Liu, "*An Adaptive Mesh Base Protocol for Geocast routing*", Journal to Parallel and Distributed Computing", Volume 63, Issue 2, February 2003.

[13]    T-F Shih and H-C Yen,"*Core Location-Aided ClusterBased Routing Protocol for Mobile Ad hoc Networks*", 10th WSEAS International Conference on Communications.

[14]    J. Zhou, "*An Any-cast based Geocasting Protocol for MANET", ISPA 2005, LNCS 3758, pp.915-926, 2005.*

[15]    Y.B.Ko and N.H.Vaidya, "*Anycasting based protocol for geocast service in mobile ad hoc networks*", Computer Networks Journals, vol.41, no. 6, 2003.

[16]    A.K.Sadek, W.Yu and K.J.Ray Liu, "*On the Energy Efficiency of Cooperative Communications in Wireless Sensor Networks"*, ACM transactions on Sensor Networks, vol. 6, N0.1: December 2009.

[17]    Y.Dong, W-K Hon, D.K.Y. Yau, J-C Chin, "*Distance Reduction in Mobile Wireless Communication;Lower Bound Analysis & practical Attainment",*IEEE transactions on Mobile Computing, vol. 8 no. 2, February 2009.

[18]    M. Zorzi and R R Rao, "*Geographic Random Forwarding(GeRaF) for Adhoc & Sensor Networks:Energy  & Latency Performance",*IEEE transactions on Mobile Computing, vol.2 no.4, October-December 2003.

[19]    A.G. Ruzzelli, G.'Hare and R. Higgs, "*Directed Broadcast eith Overhearing for Sensor Networks*",     ACM Transactions on Sensor Networks, Vol.6, No.1, December 2009.

## Authors

**Kaushik Ghosh** is an Assistant Professor in the Department of Computer Science & Engineering, Mody Institute of Technology and Science, Lakshmangargh(Sikar), India. He received his Bachelor of Engineering degree in Electrical & Electronics Engineering from Sikkim Manipal Institute of Technology and has done his M.Tech in Computer Technology from Jadavpur University, India. His area of research includes Mobile Adhoc Networks (MANETs) and Sensor Networks.

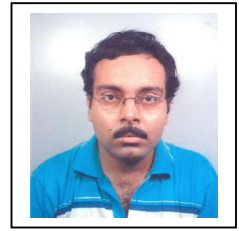

**Sarbani Roy** received the PhD degree in Engineering from Jadavpur University, Kolkata, India in July, 2008, and the M-Tech degree in Computer Science & Engineering, the MSc degree in Computer and Information Science and BSc Honors degree in Computer Science from University of Calcutta, Kolkata, India. Since 2006 she has been a faculty of the Department of Computer Science and Engineering, Jadavpur University, Kolkata, India. Her research interests are in the area of High Performance Computing, Grid Computing,Distributed Computing, Wireless Sensor Network and Mobile Computing

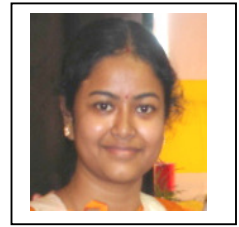

**Pradip K Das** is a Professor and Dean in the Faculty of Engineering & Technology, Mody Institute of Technology & Science, Lakshmangargh(Sikar), India. Earlier, he was a Professor and Head in the Department of Computer Science & Engineering and the founder Director of the School of Mobile Computing & Communication, Jadavpur University, India. He received is B.E. and M.E. degrees in Electronics and Ph.D (Engg.) in Computer Engineering from Jadavpur University.

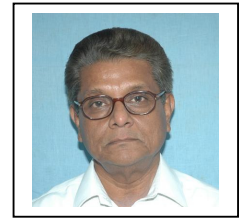